\documentclass[aps,prl,preprint,amsmath,amssymb,amsfonts]{revtex4-2}

\usepackage{graphicx}
\graphicspath{{./}}

\usepackage{hyperref}%

\usepackage{siunitx}
\newcommand{\rxx}{$R_{xx}$}
\newcommand{\sxx}{$\sigma_{xx}$}
\newcommand{\vaexx}{$V^{\mathrm{AE}}_{xx}$}
\newcommand{\dsxx}{$\Delta\sigma_{xx}$}
\newcommand{\dvaexx}{$\Delta V^{\mathrm{AE}}_{xx}$}
\newcommand{\LNO}{LiNbO\textsubscript{3}}
\newcommand{\ee}{\mathrm{e}}
\newcommand{\ii}{\mathrm{i}}

\begin{document}

\title{Acoustoelectric Probing of Fractal Energy Spectra in Graphene/hBN Moir\'e Superlattices}

\author{Wenqing Song\textsuperscript{1,2}$^\dagger$}
\author{Yicheng Mou\textsuperscript{1,2}$^\dagger$}
\author{Qing Lan\textsuperscript{1,2}}
\author{Guorui Zhao\textsuperscript{1,2}}
\author{Zejing Guo\textsuperscript{1,2}}
\author{Jiaqi Liu\textsuperscript{1,2}}
\author{Tuoyu Zhao\textsuperscript{1,2}}
\author{Cheng Zhang\textsuperscript{1,2}$^\ast$}
\author{Wu Shi\textsuperscript{1,2}$^\ast$}

\affiliation{\textsuperscript{1}State Key Laboratory of Surface Physics and Institute for Nanoelectronic Devices and Quantum Computing, Fudan University, Shanghai 200433, China}
\affiliation{\textsuperscript{2}Zhangjiang Fudan International Innovation Center, Fudan University, Shanghai 201210, China}
\affiliation{$^\dagger$These authors contributed equally to this work.}
\affiliation{$^\ast$Corresponding authors. E-mail: zhangcheng@fudan.edu.cn (C. Z.); shiwu@fudan.edu.cn (W. S.).}


\begin{abstract}
Moir\'e superlattices with long-range periodicity exhibit Hofstadter energy spectra under accessible magnetic fields, enabling the exploration of emergent quantum phenomena through a hierarchy of fractal states.
However, higher-order features, located at elevated energies with narrow bandwidths, typically require high carrier densities and remain difficult to resolve using conventional electrical transport due to limited sensitivity and strong background conductivity.
Here, we utilize acoustoelectric (AE) transport to probe high-order fractal states and the Hofstadter spectrum in graphene/hBN moir\'e superlattices.
Surface acoustic waves on a ferroelectric \LNO\ substrate generate an AE voltage proportional to the derivative of electrical conductivity, significantly enhancing sensitivity to weak spectral features.
Combined with substrate-induced high electron doping, this technique resolves fractal Brown-Zak oscillations up to the fifth-order and provides the first AE observation of the Hofstadter butterfly, revealing high-order fractal magnetic Bloch states and symmetry-broken Landau levels over a wide carrier density range.
Our results establish AE transport as a powerful derivative-sensitive probe for emergent fractal quantum states in moir\'e-engineered 2D systems.
\end{abstract}

\maketitle

\section{Main text}\label{sec1}

The interplay between a perpendicular magnetic field and a two-dimensional lattice periodicity reconstructs the electronic spectrum into a self-similar fractal known as the Hofstadter butterfly \cite{hofstadterEnergyLevelsWave1976}. 
At rational magnetic flux values ($\mathit{\Phi}/\mathit{\Phi}_0 = p/q$, where $\mathit{\Phi} = B\cdot A$ is the flux per unit cell, $\mathit{\Phi}_0 = h /e$ is the flux quantum, and $p$, $q$ are coprime integers), energy gaps recur at increasingly finer scales.
In natural atomic crystals, resolving this fractal structure requires prohibitively large magnetic fields exceeding \qty{e3}{T}.
The advent of artificial superlattices, especially moir\'e superlattices formed in van der Waals heterostructures, has drastically reduced the required fields, enabling experimental access to Hofstadter physics under realistic conditions \cite{albrechtEvidenceHofstadtersFractal2001,ponomarenkoCloningDiracFermions2013, deanHofstadtersButterflyFractal2013a}. 
These moir\'e systems offer long-range periodicity, multiband coupling, and diverse lattice geometries, providing a versatile platform for exploring emergent fractal and topological phenomena \cite{deanHofstadtersButterflyFractal2013a, krishnakumarHightemperatureQuantumOscillations2017b,serlinIntrinsicQuantizedAnomalous2020b,caiSignaturesFractionalQuantum2023,suMoiredrivenTopologicalElectronic2025}.
However, despite numerous transport studies reported to date, resolving the fine structure of the Hofstadter spectrum, especially high-order fractal gaps and associated topological transitions, remains a major experimental challenge using conventional transport measurements \cite{krishnakumarHighorderFractalStates2018a,shiHighorderFractalQuantum2023}.

To address this, derivative-sensitive probes such as scanning probe microscopy \cite{yuCorrelatedHofstadterSpectrum2022a, nuckollsSpectroscopyFractalHofstadter2025} and thermopower measurements \cite{sultanaDetectionFractionalQuantum2025,ghoshThermopowerProbesEmergent2025} have been used to enhance the visibility of spectral features and interaction-driven states.
Recently the interaction of surface acoustic waves (SAWs) and two-dimensional (2D) materials has attracted much interest \cite{preciadoScalableFabricationHybrid2015,satzingerQuantumControlSurface2018,yokouchiCreationMagneticSkyrmions2020,yokoiNegativeResistanceState2020,guAcoustodragPhotovoltaicEffect2024}.
Acoustoelectric (AE) transport has emerged as a powerful alternative probe \cite{kalameitsevValleyAcoustoelectricEffect2019,sukhachovAcoustogalvanicEffectDirac2020,sonowalAcoustoelectricEffectTwodimensional2020,friessAcoustoelectricStudyMicrowaveinduced2020,dattaSpatiotemporallyControlledRoomtemperature2022,pengLongrangeTransport2D2022, mouGatetunableQuantumAcoustoelectric2024,mouCoherentDetectionOscillating2025, wangAnomalousAcoustocurrentQuantum2025}. 
In AE measurements, 2D electronic devices are fabricated on piezoelectric substrates (e.g., \LNO), where propagating SAW is coupled to charge carriers via piezoelectric and strain fields \cite{wixforthQuantumOscillationsSurfaceacousticwave1986,wixforthSurfaceAcousticWaves1989,efrosQuantizationAcoustoelectricCurrent1990,falkoAcoustoelectricDragEffect1993,esslingerUltrasonicApproachInteger1994,shiltonExperimentalStudyAcoustoelectric1995,miseikisAcousticallyInducedCurrent2012}.
The resulting momentum transfer induces directed charge motion, generating an AE current.
Unlike conventional electrical transport, AE transport probes the variation of conductivity $\sigma$, effectively measuring $\partial \sigma/\partial n$, where $n$ is the carrier density.
This derivative sensitivity amplifies weak oscillatory features and enables direct detection of high-order quantum oscillations, which are often buried under the metallic background in resistivity measurements \cite{mouGatetunableQuantumAcoustoelectric2024}.

In this Letter, we use AE transport to resolve the high-order fractal Brown-Zak (BZ) oscillations in an aligned graphene (Gr)/hexagonal boron nitride (hBN) moir\'e superlattice device on \LNO\ substrate and explore subtle fractal features in the Hofstadter butterfly spectrum.
The substrate-induced electron doping at low temperatures allows access to the high-carrier-density regimes without degrading the device quality \cite{fangQuantumOscillationsGraphene2023, mouUnexpectedLargeElectrostatic2025}, facilitating the detection of fractal quantum oscillations.
By leveraging the enhanced spectral sensitivity of AE transport, we observe fractal BZ oscillations up to the fifth order and obtain the first AE transport observation of the Hofstadter butterfly spectrum, resolving high-order magnetic Bloch states and symmetry-broken Landau levels.
These results demonstrate AE transport as a uniquely capable probe of fractal energy spectra and emergent quantum states in moir\'e-engineered 2D systems.

\begin{figure}[!t]
    \centering
    \includegraphics[]{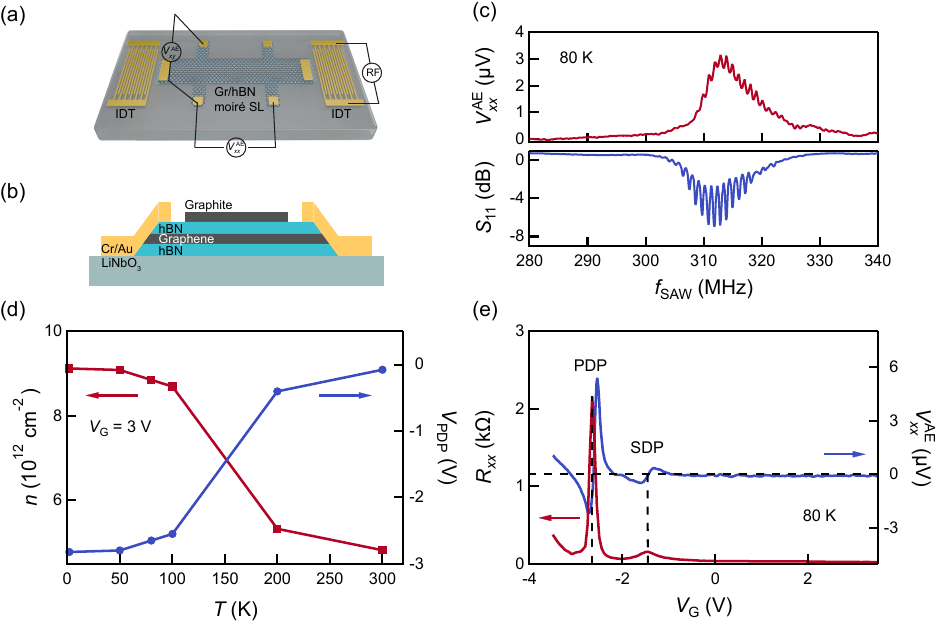}
    \caption{Integration of Gr/hBN moir\'e superlattice with SAWs on a \LNO\ substrate.
    (a) Schematic of the device with IDTs and AE measurement configuration.
    (b) Cross-sectional schematic of the device with the graphite top gate.
    (c) Frequency dependence of AE voltage \vaexx\ (top) and the IDT reflection spectrum (bottom), with a peak at $f_\mathrm{SAW} = \qty{314}{MHz}$.
    (d) Carrier density $n$ at $V_\mathrm{G} = \qty{3}{V}$ and primary Dirac-point position $V_\mathrm{PDP}$ versus temperature.
    (e) Comparison of \rxx\ and \vaexx\ versus $V_\mathrm{G}$ at \qty{80}{K}.
    Vertical dashed lines denote the primary Dirac and secondary Dirac points.}
    \label{fig1}
\end{figure}

To perform both acoustoelectric and electrical transport measurements, we fabricated an hBN-encapsulated monolayer graphene (MLG) device on a \LNO\ substrate.
Figure \ref{fig1}(a) shows the schematic of the AE transport measurement setup with delay-line-type interdigital transducers (IDTs) on a \LNO\ substrate.
Figure \ref{fig1}(b) shows the cross-sectional schematic of the device.
Further fabrication details are provided in the Methods section of Supplemental Material \cite{supplementalMaterial}.
Figure \ref{fig1}(c) shows the frequency-dependent longitudinal AE voltage \vaexx\ (top panel) and the reflection spectrum $S_{11}$ (bottom panel).
Both \vaexx\ and $S_{11}$-parameter reach maximum/minimum near the frequency $f_\mathrm{SAW} = \qty{314}{MHz}$, which is determined by the IDT geometry.
Subsequent AE transport measurements were performed at this center frequency.
Unless otherwise specified, the input RF signal power to the IDTs was \qty{3}{dBm} (\qty{2}{mW}).

The \LNO\ substrate introduces substantial electron doping in the Gr/hBN superlattices at low temperatures.
We provide the transfer curve \rxx\ versus $V_\mathrm{G}$ of the device at various temperatures in Supplemental Fig. S1 at \cite{supplementalMaterial}.
At \qty{300}{K}, it exhibits a sharp primary Dirac point (PDP) peak and two broader secondary Dirac points (SDPs) peaks, reflecting moir\'e-induced band
folding\cite{ponomarenkoCloningDiracFermions2013,moonElectronicPropertiesGraphene2014b}.
From the PDP-SDP spacing, the full-filling carrier density of the moir\'e superlattice is $4n_{0}=\qty{1.82e12}{cm^{-2}}$ with $n_{0} = 1 / S = 2/(\sqrt{3}a_\mathrm{s}^{2})$, giving an estimated moir\'e superlattice $a_\mathrm{s} = \qty{15.5}{nm}$.
Upon cooling, the transfer curve left shifts, indicating increasing \textit{n}-type doping attributed to enhanced ferroelectric polarization of the \ang{128}-Y-cut \LNO\ at low temperatures\cite{fangQuantumOscillationsGraphene2023, mouGatetunableQuantumAcoustoelectric2024,mouUnexpectedLargeElectrostatic2025}.
Figure \ref{fig1}(d) presents how the PDP position and $n$ at $V_\mathrm{G} = \qty{3}{V}$ evolve with the temperature.
The carrier density increases with decreasing temperature and gradually saturates below $\sim$ \qty{50}{K}.
Combining the substrate-induced and gate-induced doping enables carrier densities up to \qty{1e13}{cm^{-2}}, sufficient to resolve high-order fractal quantum oscillations.
Notably, mobility and moir\'e characteristics remain intact, as the PDP-SDP spacing is temperature independent.

Similar to conventional electrical transport, AE transport can reliably reflect the characteristics of the moir\'e superlattice.
Figure \ref{fig1}(e) compares \rxx\ and \vaexx\ as functions of $V_{\mathrm{G}}$. Near PDP and SDP positions, \rxx\ reaches a peak while \vaexx\ diverges and changes sign, clearly resembling the moir\'e superlattice behavior.
The AE effect originates from the interaction of SAW with 2D electron system.
As the SAW travels along the surface of a piezoelectric substrate, it generates a copropagating piezoelectric field $E_{\mathrm{p}} = E_{\mathrm{p0}}\ee^{\ii(qx-\omega t)}$, which drives a spatiotemporal-dependent current in the material \cite{rotterGiantAcoustoelectricEffect1998,bandhuMacroscopicAcoustoelectricCharge2013,zhaoAcousticallyInducedGiant2022},
\begin{equation}
    j^{\mathrm{AE}} = \frac{\sigma_{xx}\left(\omega\right)E_{\mathrm{p}}}{1+\ii\sigma_{xx}\left(\omega\right)/\sigma_{\mathrm{m}}}
    \label{eq:current1}
\end{equation}
where $\sigma_{\mathrm{m}} = v_{\mathrm{SAW}}\epsilon_0\left(1+\epsilon_\mathrm{r}\right)$ is a substrate-dependent constant, $\sigma_{xx}\left(\omega\right)$ is the frequency-dependent conductivity of the 2D electron system, and $\omega$ is the frequency of the SAW.
To screen this oscillating piezoelectric field, the electron system forms a spatially modulated charge density, leading to a nonlinear interaction with SAW.
This nonlinear coupling gives rise to a rectified d.c. current component as given by
\begin{equation}
    j^{\mathrm{AE}}_{\mathrm{d.c.}} = \frac{\mathit{\Gamma} I_{\mathrm{SAW}}}{ev_{\mathrm{SAW}}}\left.\frac{\partial \sigma_{xx}}{\partial n}\right|_{\varepsilon_{\mathrm{F}}}
    \label{eq:jdc}
\end{equation}
where $\mathit{\Gamma} \propto \sigma_{xx}\sigma_{\mathrm{m}}/\left(\sigma_{\mathrm{m}}^2+\sigma_{xx}^2\right)$ is the attenuation coefficient of the SAW, $I_{\mathrm{SAW}}$ and $v_{\mathrm{SAW}}$ are the intensity and velocity of the SAW, respectively, $e$ is the elementary charge, and $\varepsilon_\mathrm{F}$ is the Fermi energy.
The formula derivation is provided in Supplemental Material \cite{supplementalMaterial}. 
This relation directly links the AE response to the derivative of conductivity with respect to the carrier density at the Fermi level, $\partial \sigma_{xx}/\partial n$, which diverges near charge neutrality points due to the vanishing density of states (DOS).
In Gr/hBN superlattices, these divergences reflect mini-gap formation at PDP and SDPs, where the band structure is reconstructed by the superlattice potential \cite{ponomarenkoCloningDiracFermions2013}.
Unlike conventional transport, the derivative $\partial \sigma_{xx}/\partial n$ enhances sensitivity to subtle conductivity variations, enabling AE measurements to resolve higher-order fractal gaps in Hofstadter spectra that are otherwise obscured in conductivity measurements.

\begin{figure}[!t]
    \centering
    \includegraphics[]{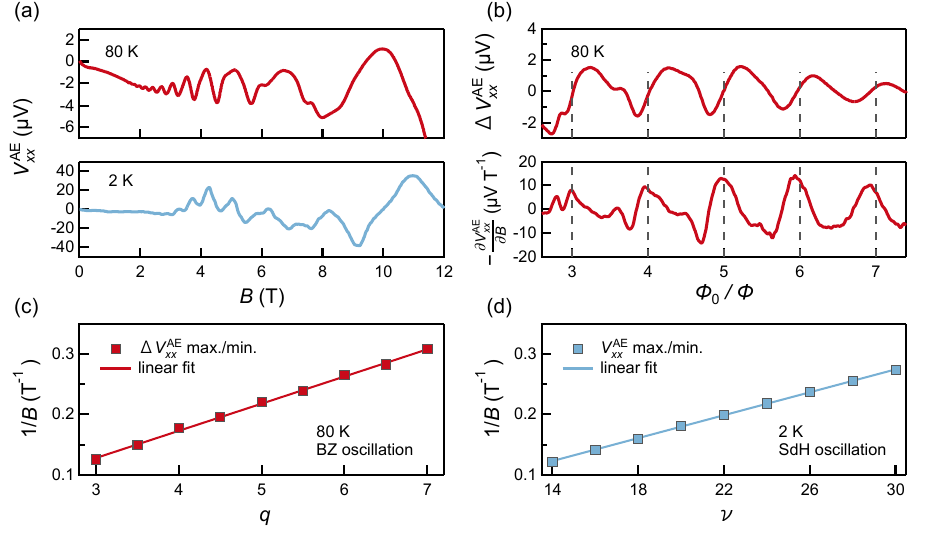}
    \caption{Quantum oscillations in Gr/hBN moir\'e superlattice measured by AE transport.
    (a) \vaexx\ versus magnetic field $B$ at \qty{80}{K} and \qty{2}{K} at fixed carrier density $n = 5.4 n_{0} = \qty{2.46e12}{cm^{-2}}$.
    (b) Oscillatory component \dvaexx\ and its field derivative $- \partial V_{xx}^\mathrm{AE}/\partial B$ plotted as functions of $\mathit{\Phi}_{0}/\mathit{\Phi}$, showing clear BZ oscillations.
    (c) Positions of maxima and minima in \dvaexx, plotted in $1/B$ versus $q$ for the primary BZ oscillations at $\mathit{\Phi}/\mathit{\Phi}_{0} = 1/q$ (\qty{80}{K}).
    (d) Positions of maxima and minima in \vaexx, plotted in $1/B$ versus LL filling factor $\nu$ from SdH oscillations at \qty{2}{K}.}
    \label{fig2}
\end{figure}

We then use AE transport to probe the quantum oscillations in Gr/hBN moir\'e superlattices.
Figure \ref{fig2}(a) shows the longitudinal AE voltage \vaexx\ as a function of the perpendicular magnetic field $B$ at \qty{80}{K} (top panel) and \qty{2}{K} (bottom panel).
The carrier density was kept at $n=5.4n_{0}=\qty{2.46e12}{cm^{-2}}$.
Clear quantum oscillations are observed.
At \qty{80}{K}, the oscillations are mainly BZ oscillations, which originate from the periodic emergence of delocalized Bloch states at commensurate magnetic fields \cite{brownBlochElectronsUniform1964a,zakMagneticTranslationGroup1964a,krishnakumarHightemperatureQuantumOscillations2017b}.
The frequency of BZ oscillations is determined by the superlattice periodicity, independent on the carrier density.
The top panel of Fig. \ref{fig2}(b) exhibits the oscillatory component \dvaexx\ as a function of $\mathit{\Phi}_{0}/\mathit{\Phi}$ at \qty{80}{K}, which was obtained by subtracting a smooth background from \vaexx.
Comparison with the electrical conductivity oscillations is shown in Supplemental Fig. S2 at \cite{supplementalMaterial}. 
A phase difference is found between conductivity and AE measurements, attributing to the second-order effect of SAW producing AE transport.
According to Eq. (\ref{eq:jdc}), \dvaexx\ is directly proportional to the first derivative of \dsxx, accounting for the observed phase difference.
The bottom panel of Fig. \ref{fig2}(b) shows $-\partial V^{\mathrm{AE}}_{xx}/\partial B$ as a function of $\mathit{\Phi}_{0}/\mathit{\Phi}$, resembling the BZ oscillations.
Due to the relatively low carrier density ($n/n_{0} = 5.4$), only primary BZ oscillation peaks ($\mathit{\Phi}/\mathit{\Phi}_{0} = 1/q$) are visible in \dsxx, while \dvaexx\ exhibits additional peaks between the primary ones, corresponding to the fractal Bloch states at $\mathit{\Phi}/\mathit{\Phi}_{0} = 2 / q$.
Figure \ref{fig2}(c) shows the positions of the maxima and minima of \dvaexx associated with these primary BZ oscillations, plotted as a function of $q$.
Linear fit of the data yields oscillation period $B_{0} = \qty[uncertainty-mode=separate]{22.2(0.3)}{T}$.
From $B_{0}$, we can calculate the moir\'e superlattice periodicity via $\mathit{\Phi}_{0} = B_{0}\sqrt{3}a_{\mathrm{s}}^2/2$, obtaining $a_{\mathrm{s}} = \qty{14.7}{nm}$, consistent with the periodicity estimated from the full-filling carrier density (see Fig. \ref{fig1}(e)).

In addition to BZ oscillations, Shubnikov--de Haas (SdH) oscillations arise at \qty{2}{K} due to the cyclic filling of Landau levels (LLs), as shown in the bottom panel of Fig. \ref{fig2}(a).
The carrier density filled in the LLs is given by $n = \nu B/\mathit{\Phi}_0$ \cite{zhangExperimentalObservationQuantum2005b}, where $\nu$ is the LL filling factor. Consequently, the carrier density of the Gr/hBN moir\'e superlattice can be extracted from the SdH oscillations.
The positions of local maxima and minima of SdH oscillations in $1/B$ are plotted in Fig. \ref{fig2}(d), resulting in the carrier density of $n = \qty{2.50e12}{cm^{-2}}$, in good agreement with the $n=\qty{2.46e12}{cm^{-2}}$ estimated from field effect doping.
These results prove that AE transport can serve as a reliable tool for characterizing the electronic properties of moir\'e superlattices.

\begin{figure}[!t]
    \centering
    \includegraphics[]{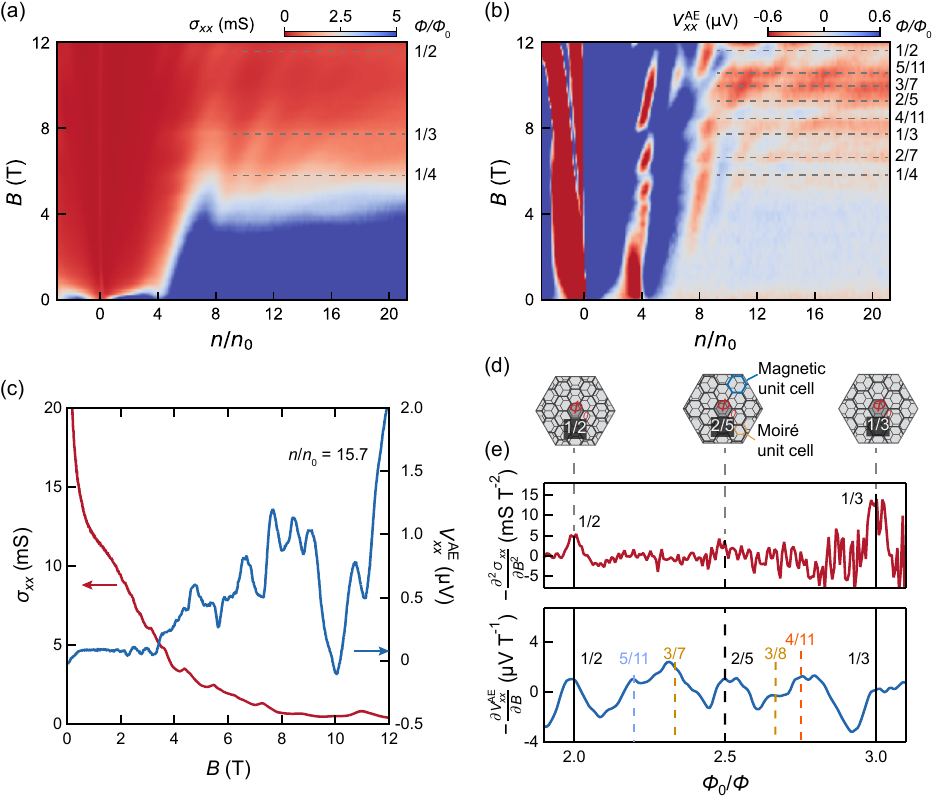}
    \caption{High-order fractal BZ oscillations in Gr/hBN moir\'e superlattice resolved via electrical and AE transport.
    (a-b) \sxx\ and \vaexx\ versus normalized carrier density $n/n_{0}$ and $B$ at \qty{80}{K}.
    Horizontal dashed lines mark magnetic Bloch states at flux values $\mathit{\Phi}/\mathit{\Phi}_{0} = p/q$.
    (c) Vertical line cuts of (a) and (b) at $n = 15.7 n_{0} = \qty{7.15e12}{cm^{-2}}$, showing \sxx\ and \vaexx\ versus $B$.
    (d) Schematic of magnetic and moir\'e unit cells at $\mathit{\Phi}/\mathit{\Phi}_{0} = p/q$.
    (e) Second derivative of \sxx\ and first derivative of \vaexx\ versus $\mathit{\Phi}_{0}/\mathit{\Phi}$ at $n = 15.7 n_{0}$.
    Vertical dashed lines mark high-order fractal BZ oscillations, resolved in AE transport up to the fifth order.}
    \label{fig3}
\end{figure}

More importantly, AE transport enables the detection of subtle features in high-order quantum oscillations and fractal energy spectra that are difficult to resolve by conventional electrical transport.
Figures \ref{fig3}(a) and \ref{fig3}(b) display \sxx\ and \vaexx\ as functions of $B$ and the normalized carrier density $n/n_{0}$ at \qty{80}{K}, respectively.
Figure \ref{fig3}(c) presents vertical line cuts at $n = 15.7 n_0 = \qty{7.15e12}{cm^{-2}}$, showing the original data of \sxx\ and \vaexx\ versus $B$.
At the high temperature of \qty{80}{K}, SdH oscillations are greatly suppressed in \sxx, yet faint Landau-fan trajectories originating from $n/n_{0} = 0$ and $4$ remain discernible in \vaexx.
We focus on the high magnetic field and high carrier density regime, where high-order BZ oscillations are expected \cite{krishnakumarHighorderFractalStates2018a,shiHighorderFractalQuantum2023}.
Above \qty{4}{T}, horizontally invariant features appear in \vaexx\ [dashed lines in Fig. \ref{fig3}(b)], corresponding to commensurate flux ratios $\mathit{\Phi}/\mathit{\Phi}_0 = p/q$ and indicating the formation of delocalized fractal magnetic Bloch states.
Figure \ref{fig3}(d) schematically illustrates the magnetic and moir\'e unit cells, where $q$ moir\'e unit cells enclose $p$ magnetic flux quanta at $\mathit{\Phi}/\mathit{\Phi}_0 = p/q$ \cite{hofstadterEnergyLevelsWave1976,nuckollsSpectroscopyFractalHofstadter2025}.
In contrast, \sxx\ exhibits local maxima only at the primary BZ states, as shown in Fig. \ref{fig3}(a).
Moreover, as $B$ increases, conventional transport becomes
dominated by strong magnetoresistance (Fig. \ref{fig3}(c)), masking high-order oscillations.
However, the amplitude of BZ oscillations in \vaexx\ grows stronger with increasing $B$, independent of background conductivity due to its derivative nature.
To be more quantitative, we further plot the second derivative $-\partial^2 \sigma_{xx} / \partial B^2$ following established procedures \cite{krishnakumarHighorderFractalStates2018a} and also the first derivative $-\partial V_{xx}^\mathrm{AE} / \partial B$ as functions of $\mathit{\Phi}_0 / \mathit{\Phi}$ in the upper and lower panels, respectively, of Fig. \ref{fig3}(e).
While higher-order features remain indiscernible with large noise in \sxx\ amplified by numerical differentiation, AE transport yields a markedly higher signal-to-noise ratio and reveals additional fractal oscillations up to the fifth
order.
This direct comparison confirms the superior sensitivity of AE transport in resolving high-order fractal states obscured by strong background conductivity.

\begin{figure}[!t]
    \centering
    \includegraphics[]{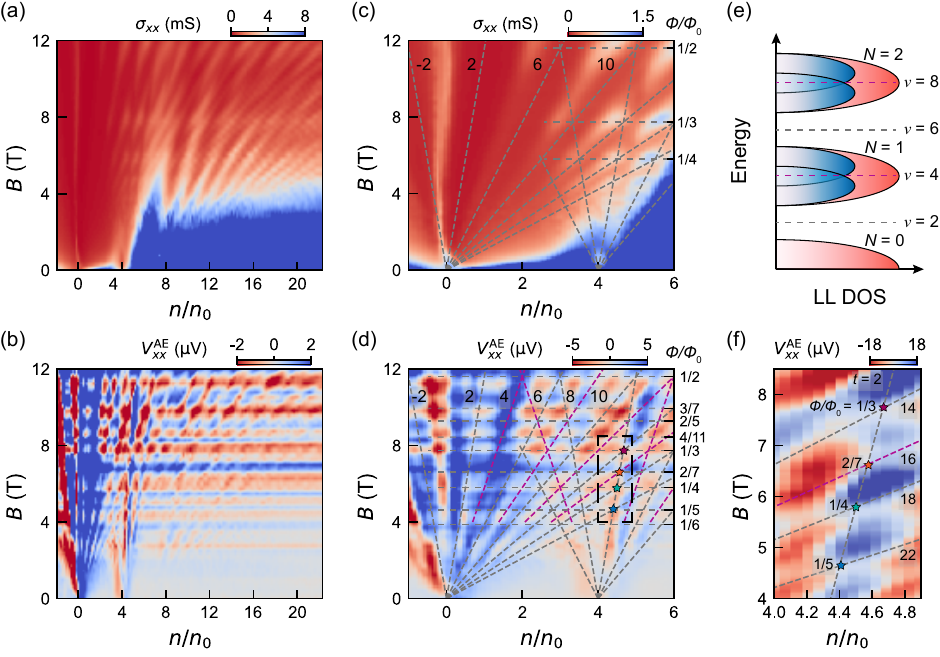}
    \caption{Electrical and AE transport observation of Hofstadter butterfly in Gr/hBN moir\'e superlattice.
    (a-b) \sxx\ and \vaexx\ versus $n/n_{0}$ and $B$ at \qty{3.5}{K}.
    (c-d) Zoomed-in views of (a) and (b) highlighting linear Landau-fan trajectories (sloped dashed lines), fractal states at commensurate fields (horizontal dashed lines), and symmetry-broken Landau fans (purple dashed lines).
    (e) Schematic of the symmetry-broken LLs in graphene.
    Red areas represent =fourfold-degenerate LLs, and blue areas indicate partially
    symmetry-broken LLs.
    (f) High-resolution AE map of $V_{xx}^\mathrm{AE}\left( n/n_{0}, B \right)$ within the dashed rectangle region in (d), resolving intersections of primary and secondary Landau fans at commensurate flux values $\mathit{\Phi}/\mathit{\Phi}_{0} = 1/3$, $2/7$, $1/4$ and $1/5$ (marked as stars).
    RF power: \qty{0}{dBm} in (b) and (d), \qty{3}{dBm} in (f).}
    \label{fig4}
\end{figure}

We further report the first AE transport observation of Hofstadter butterfly in Gr/hBN moir\'e superlattice, resolving its fractal energy spectrum across ultrahigh carrier densities.
Figures \ref{fig4}(a) and \ref{fig4}(b) present \sxx\ and \vaexx\ as functions of $n/n_{0}$ and $B$ measured at \qty{3.5}{K} in the same superlattice device, covering densities up to
\qty{1e13}{cm^{-2}}.
By tracing the minima of \sxx\ and sign-reversing points of \vaexx, we identify Landau-fan trajectories following the linear Diophantine relation $n/n_0 = t\mathit{\Phi}/\mathit{\Phi}_0 + s$, where $t$ and $s$ are integers and represent the Landau gap indices and Bloch bandfilling, respectively (see Supplemental Fig. S3 at \cite{supplementalMaterial}).
Both datasets show Landau quantization as marked in the zoomed-in maps [Figs. \ref{fig4}(c) and \ref{fig4}(d)], but symmetry-broken LLs [purple dashed lines in Fig. 4(d)] emerge only in \vaexx, where the four-fold degeneracy of graphene is partially lifted as illustrated in Fig. 4(e).
These features are broadened beyond recognition in \sxx\ [Fig. \ref{fig4}(c)], because of imperfect device quality and limited sensitivity of electrical transport, but appear clearly in AE transport due to its enhanced spectral resolution.

Additional horizontal features at commensurate magnetic fields [Figs. \ref{fig4}(b) and \ref{fig4}(d)] mark the formation of high-order fractal states (up to $p$ = 4), which are invisible in \sxx\ within the same device and conditions [Figs. \ref{fig4}(a) and \ref{fig4}(c)].
The resulting tilted checkerboard pattern in AE map [Fig. \ref{fig4}(d)] reproduces the expected Hofstadter butterfly, providing additional fractal spectra details that are obscured in the conductivity map.
The horizontal structures in AE map reflect transitions between
localized and delocalized Bloch states, revealing the underlying fractal topology of the Hofstadter spectrum.
At incommensurate $B$, electrons form localized cyclotron orbits, while at commensurate $B$ they become delocalized across the moir\'e lattice.
These crossovers give rise to the complex spectral hierarchy.
To resolve this more clearly, we performed a high-resolution AE scan in the range of $4.0 < n/n_{0} < 4.9$ and $\qty{4.0}{T} < B < \qty{8.5}{T}$, producing a well-defined tilted checkerboard pattern in \vaexx\ and resolving intersections at $\mathit{\Phi}/\mathit{\Phi}_{0} = 1/3$, $2/7$, $1/4$ and $1/5$.
Across each fan trajectory, \vaexx\ reverses sign, consistent with effective changes in carrier type, but remains continuous at the intersection points [stars in Fig. \ref{fig4}(f)], where
localized and delocalized states merge.
Together, these observations demonstrate that AE transport directly visualizes the fractal hierarchy of Hofstadter butterfly and captures quantum phase transitions with higher resolution.

We now discuss why AE transport provides enhanced sensitivity for revealing fractal quantum oscillations.
Higher-order BZ oscillations arise from the commensurability between magnetic flux and the moir\'e lattice periodicity.
These higher-order states typically appear at elevated energies, with narrow bandwidths and reduced spectral weights \cite{moonElectronicPropertiesGraphene2014b, krishnakumarHighorderFractalStates2018a}, requiring high carrier densities to raise the Fermi level into relevant minibands.
However, in this high-doping regime, conventional electrical transport becomes increasingly insensitive to weak oscillatory features.
The overall conductivity is dominated by a metallic background that masks fine spectral structure, while enhanced screening and suppressed energy-dependent scattering at high densities \cite{terekhovScreeningCoulombImpurities2008} further reduce the visibility of quantum oscillations in conventional electrical transport.
In contrast, AE transport is governed by the derivative of the conductivity at the Fermi level.
This differential sensitivity makes AE measurements inherently more responsive to subtle variations in the electronic structure.
As a result, AE transport can detect weak spectral features that may be otherwise obscured by background conductivity in conventional electrical measurements.
We further compare AE with other techniques, such as STM and thermopower measurements, in Supplemental Material at \cite{supplementalMaterial}.
Overall, AE transport provides a simple bulk probe that is fully compatible with standard electrical device architectures and inherently derivative sensitive, offering a broadly accessible platform for exploring correlated quantum phenomena in diverse 2D moir\'e systems.

Physically, a fractal Bloch state with $\mathit{\Phi}/\mathit{\Phi}_{0} = p/q$
corresponds to $q$ moir\'e unit cells enclosing $p$ flux quanta.
For example, $\mathit{\Phi}/\mathit{\Phi}_{0} = 4/11$ spans eleven moir\'e cells, giving to a characteristic length scale of $\sim$ \qty{100}{nm}.
SAWs couple efficiently with long-range periodic modulations, including moir\'e superlattices and charge density waves\cite{rudolphLongrangeExcitonTransport2007,friessNegativePermittivityBubble2017, fangQuantumOscillationsGraphene2023}.
As the SAW wavelength (\qty{12}{\um}) used here is much larger than the moir\'e period, our experiments operate in the long-wavelength limit, where SAW acts as a spatially uniform perturbation averaged over many moir\'e cells, without commensurability or resonance effects.
In this work, we employed only delay-line-type IDTs for AE transport measurements.
Alternative resonant IDTs with close spacing, commonly used in SAW attenuation studies, could further enhance coupling and enable fully non-contact quantum transport probes\cite{fangQuantumOscillationsGraphene2023,wuProbingQuantumPhases2024, caoPhaseDiagramMapping2025}.
These SAW-based approaches offer valuable complementary capabilities to conventional transport techniques, thereby broadening the investigative scope of quantum materials.

\section{Conclusion}

In summary, we show that AE transport is an effective probe of high-order quantum oscillations and the Hofstadter butterfly in Gr/hBN moir\'e superlattices.
Its intrinsic sensitivity to the derivative of the conductivity enables clear detection of subtle fractal features, including high-order magnetic Bloch states and symmetry-broken Hofstadter energy spectra that are often challenging to resolve via conventional electrical transport.
The integration of \LNO-induced electron doping with the use of SAW-based techniques establishes a powerful platform for exploring emergent quantum phenomena in moir\'e-engineered 2D systems with enhanced sensitivity.

\section{Acknowledgments}
\begin{acknowledgments}
We gratefully thank Haiwen Liu, Yijia Wu and Hua Jiang for helpful discussions.
This work is supported by the National Key Research and Development Program of China (Grants No. 2024YFA1409003, No. 2024YFB3614103, and No. 2022YFA1405700), the National Natural Science Foundation of China (Grants No. 12274090, No. 12574187, No. 92365104, and No. 12174069), the Shanghai Pilot Program for Basic Research-Fudan University 21TQ1400100 (25TQ001), and Shanghai QiYuan Innovation Foundation.
Part of the sample fabrication was performed at the Fudan Nanofabrication Laboratory, Fudan University.
\end{acknowledgments}

%


\newpage

\newpage

\newpage

\newpage

\end{document}